\documentstyle[12pt]{article}
\begin{document}
\baselineskip 22pt plus 2pt

{}~ \hfill TECHNION-PH-95-18 \\
\vspace{1cm}

\begin{center}
{\large \bf Long Distance Contribution to $s\rightarrow d\gamma$ and
Implications
for $\Omega^-\rightarrow \Xi ^-\gamma, \ B_s \rightarrow B_d^*\gamma$
and $b\rightarrow s\gamma$.}\\
\vspace{0.50cm}

G. Eilam$^a$, A. Ioannissian$^{a,b}$, R.R. Mendel$^{a,c}$ and P. Singer$^a$
\vspace{0.50cm}

\parbox{13cm}
{$a$ ~~Dept. of Physics, Technion -- Israel Inst. of Tech., Haifa 32000,
Israel\\
$b$ ~~On leave from Yerevan Physics Institute, Alikhanyan Br. 2, Yerevan,
\hspace*{0.50cm} 375036,
Armenia.\\
$c$ ~~On sabbatical leave from Dept. of Applied Math., Univ. of Western
\hspace*{0.50cm} Ontario,
London,  Ontario, Canada. }
\end{center}
\vspace{0.25cm}

\noindent{\bf Abstract}

We estimate the long distance (LD) contribution to the magnetic part of the
$s\rightarrow d\gamma$ transition using the Vector Meson Dominance
approximation $(V=\rho,\omega,\psi_i)$.  We find that this contribution may be
significantly larger than the short distance (SD)  contribution to
$s\rightarrow
d\gamma$ and could possibly saturate the present experimental upper bound
on the
$\Omega^-\rightarrow\Xi^-\gamma$ decay rate,
$\Gamma^{\rm MAX}_{\Omega^-\rightarrow\Xi^-\gamma}
\simeq 3.7\times10^{-9}$eV.
For the decay $B_s\rightarrow B^*_d\gamma$, which is driven by $s\rightarrow
d\gamma$ as well, we obtain an upper bound on the branching ratio
$BR(B_s\rightarrow B_d^*\gamma)<3\times10^{-8}$ from
$\Gamma^{\rm MAX}_{\Omega^-\rightarrow\Xi^-\gamma}$.

Barring the possibility that the Quantum Chromodynamics coefficient
$a_2(m_s)$ be much smaller than 1,
$\Gamma^{\rm MAX}_{\Omega^-\rightarrow\Xi^-\gamma}$ also implies the
approximate relation $\frac{2}{3} \sum_i \frac{g^2_{\psi_i}(0)}{m^2_{\psi_i}}
 \simeq
\frac{1}{2} \frac{g^2_\rho(0)}{m^2_\rho}
+ \frac{1}{6}\frac{g^2_\omega(0)}{m^2_\omega}$.  This relation agrees
quantitatively with a recent independent estimate of the l.h.s. by
Deshpande et al., confirming that the LD contributions to $b\rightarrow
s\gamma$ are small.  We find that these amount to an increase of
$(4\pm2)\%$ in the magnitude of the $b\rightarrow s \gamma$ transition
amplitude, relative to the SD contribution alone.\\

\noindent {\bf 1. \ Introduction and Overview}\\

The investigation of the quark radiative transition $b\rightarrow s\gamma$
has been an important focus of attention in recent years [1] both because
of experimental measurements [2] and because long distance (LD)
corrections to the Standard Model (SM) predictions for the short-distance
(SD) contributions  are estimated to be small [3].
(For exclusive $B\rightarrow K^*\gamma$ decays see
Ref.~[4]).  Thus, this transition constitutes an excellent laboratory
to test the SM or possible high energy deviations thereof [5].  It has
been pointed out recently [6] that for the $c\rightarrow u\gamma$
transition the situation is reversed, with the LD contributions
dominating over the SD ones by many orders of magnitude.

In this paper we investigate the analogous quark transition $s
\rightarrow d\gamma$ and two exclusive hadronic processes, $\Omega^-
\rightarrow \Xi^-\gamma$ and $B_s \rightarrow B^*_d\gamma$, where it
plays an important role.  Throughout this paper we are concerned with the
magnetic transition only, since the charge-radius one vanishes for real
photons.

The SD contribution to $s\rightarrow d\gamma$ has been investigated before
[see e.g.: 7,8,9] and we simply repeat the calculations, using updated values
for
the relevant QCD coefficients.  Applying the quark model formalism of
Ref.~[10] we find that the SD $s\rightarrow d\gamma$ contribution (by
itself) to the
$\Omega^- \rightarrow \Xi^-\gamma$ decay rate is far below (by a factor of
order 600) the present experimental upper limit [11]:
\begin{eqnarray}
\Gamma (\Omega^- \rightarrow \Xi^-\gamma)<3.7\times10^{-9} \ {\rm eV \
(90\%CL}).     
\end{eqnarray}

Hadronic LD effects that involve light mesons in loops are estimated
to be small${[8,12]}$, comparable to the SD contributions.

On the other hand, by using a Vector Meson Dominance (VMD) approximation
for the LD contribution to the $s\rightarrow d\gamma$ transition
(along the lines discussed by Deshpande et al. [3] for $b\rightarrow s\gamma$),
we find LD contribution that are likely to be significantly larger than the
SD ones.  In fact, the rate for $\Omega^-\rightarrow\Xi^-\gamma$ may not be
far from the experimental bound (1), due to this VMD contribution.  The
resulting VMD amplitude is approximately proportional to
\begin{eqnarray}
a_2(m_s)
\left[
\frac{2}{3} \sum_i \frac{g^2_{\psi_i}(0)}{m^2_{\psi_i}} -
\frac{1}{2} \frac{g^2_\rho(0)}{m^2_\rho}
- \frac{1}{6}\frac{g^2_\omega(0)}{m^2_\omega} \right]      
\end{eqnarray}
where $a_2(m_s)$ is a Quantum Chromodynamics (QCD)
 coefficient [13] and the $g_V(0)$'s are
the usual vector meson-photon couplings, evaluated at $q^2=0$.  Although a
direct estimate of $a_2(m_s)$ is not reliable because we are well into
the low energy region where perturbation theory cannot be trusted, we
can use ```smoothness'' arguments to extrapolate from the phenomenologically
determined values $a_2(m^2_b)=0.24\pm 0.04$ [3] and $a_2(m^2_c) = 0.55
\pm 0.1$ [13] to obtain $a_2(m_s) \geq O(0.5)$.
We then apply the formalism of Ref.~[10] to get an expression for the
$\Omega^-\rightarrow\Xi^-\gamma$ decay rate from our SD+VMD\linebreak
$s\rightarrow
d\gamma$ amplitude.  (Notice that there are no pole contributions to this
decay).  It turns out that if the above rough estimate $a_2(m_s)\geq O(0.5)$
is correct then the experimental limit (1) can be satisfied only if the
contribution of the $\psi_i$ resonances in the parenthesis of eq.~(2) cancels,
at a level of 30\% or better accuracy, the $\rho$ and $\omega$ meson
contributions,
which can be reliably obtained from the $\rho$ and $\omega$ leptonic widths
[14].  The limit (1) then forces the approximate relation at $q^2=0$:
\begin{eqnarray}
\frac{2}{3} \sum_i \frac{g^2_{\psi_i}(0)}{m^2_{\psi_i}}  \simeq
\frac{1}{2} \frac{g^2_\rho(0)}{m^2_\rho}
+ \frac{1}{6}\frac{g^2_\omega(0)}{m^2_\omega}       
\end{eqnarray}
which is highly nontrivial, and may be interpreted as a remnant of the
badly broken SU(4)$_F$ symmetry.

The relation (3) turns out to be very useful for the $b\rightarrow
s\gamma$ decays.  As noted in Ref.~[3], in the VMD approximation the main LD
contributions to this decay can be expressed in terms of the l.h.s. of
eq.~3.  In Ref. [3], the sum in the l.h.s. of eq.~(3) is estimated by using
measured leptonic widths of the $\psi_i$ states and $\psi$ photoproduction
data as well as an assumption about the higher $\psi$ excitations.  We
estimate this sum with better accuracy by replacing experimental values
for $g_\rho(0)$ and $g_\omega(0)$ in the r.h.s. of eq.~(3) and find very
good quantitative agreement with the central value obtained in Ref.\ [3].
We thus confirm the main result of Ref.\ [3] that LD contributions to
$b\rightarrow s\gamma$ are of order of a few percent.  According to our
explicit estimate, these corrections amount to an increase of $(4\pm 2\%)$ in
the magnitude of the $b\rightarrow s\gamma$ transition amplitude, relative to
the SD contribution alone.

Finally, we also apply the SD + VMD approximation for $s\rightarrow d\gamma$
to the unusual decay mode $B_s\rightarrow B^*_d\gamma$, where the $b$ quark
plays
the ``spectator'' role.  We point out that this decay (followed by
$B^*_d\rightarrow
B_d\gamma$) has a clear experimental signature of 2 monochromatic photons
of energies $\simeq 50$MeV each. We find, using the limit of eq.~(1), a small
but hopefully measurable
branching ratio $BR(B_s\rightarrow B_d^*\gamma)<3\times10^{-8}$.\\

\noindent{\bf 2. \ SD Contribution to the $s\rightarrow d\gamma$ Amplitude}

The SD amplitude relevant to the $s\rightarrow d\gamma$ transition can be
expressed as
\begin{eqnarray}
A_{\rm SD} = - \frac{e}{8\pi^2} \frac{G_F}{\sqrt{2}} F_2
(\mu^2)\bar{d}\sigma^{\mu\nu}
[m_sR+ m_d L]sF_{\mu\nu} \ ,     
\end{eqnarray}
where $m_s$, $m_d$ are current quark masses and $F_2(\mu^2)$ is a form
factor evaluated at a low scale $\mu \geq 0(m_s)$ which includes (dominant)
QCD corrections.  Early estimates of $F_2(\mu^2)$ [7,8,12] were in the
approximate range $0.15 -          0.36$ [15] while we obtain by explicit
calculation, using $\alpha_s(m^2_c)\simeq 0.3, \ \alpha_s(\mu^2)=0.9$
in the formulas given in Ref.\ [16], a somewhat smaller value $F_2(\mu^2)
\simeq 0.1$, which
will be used below (see also Ref.~[9]).\\

\noindent{\bf 3. \ LD Contribution to $s\rightarrow d\gamma$}

To estimate the LD contribution to $s\rightarrow d\gamma$  we use the VMD
approximation in analogy to the formalism used in Ref.\ [3] for $b\rightarrow
s\gamma$.  As an intermediate step one defines a transverse amplitude
$A(s\rightarrow dV(q))_T \ (V=\psi_i,\rho,\omega$ in this case) and then
introduces the $V$ to $\gamma$ conversion vertices, setting
$q^2=0$.  Using Gordon decomposition we find that the LD amplitude
for the $s\rightarrow d\gamma$  transition is
\begin{eqnarray}
A_{\rm LD} &=& - e \frac{G_F}{\sqrt{2}} V_{c s} V^*_{cd} a_2(\mu^2)
\left(\frac{2}{3} \sum_i \frac{g^2_{\psi_i}(0)}{m^2_{\psi_i}} -
\frac{1}{2} \frac{g^2_\rho(0)}{m^2_\rho}
- \frac{1}{6}\frac{g^2_\omega(0)}{m^2_\omega}\right)      \nonumber \\
&&\cdot \frac{1}{M_s^2-M^2_d} \bar{d}\sigma^{\mu\nu}
[M_sR-M_dL]s F_{\mu\nu} \ ,      
\end{eqnarray}
where we have used $V_{cs}V^*_{cd} \simeq - V_{us}V^*_{ud}, \ a_2(\mu^2)$
is a QCD coefficient the value of which is taken from phenomenology
in the context of the factorization approximation [13], and the $g_V(q^2)$
factors are defined in the usual way, e.g. $\langle\psi(q)|\bar{c}\gamma_\mu
c| 0\rangle = ig_\psi(q^2)\epsilon^+_\mu(q)$.  We have not included
possible contributions from the $\rho$ and $\omega$ radial excitations
$(\rho^\prime, \rho^{\prime\prime}, \dots, \omega^\prime,
\omega^{\prime\prime},
\dots$) because we think that their contribution is much smaller and is already
taken into account to a significant degree in the SD amplitude (4).
The $\psi$ excitations should be included however, because they are
narrow resonances that are clearly distinguished from the $c\bar{c}$
continuum.  Note that due to the hadronic nature of the VMD approximation,
$M_s$ and $M_d$ should correspond to ``constituent'' mass parameters.
(The use of ``constituent quark'' spinors in deriving (5) should take into
account to some extent non-perturbative effects such as chiral
symmetry breaking and confinement).  In any case, it turns out that
only the combination
$\frac{\sqrt{M_s^2+M_d^2}}{M^2_s-M^2_d}$ which has a similar magnitude for
``constituent'' or ``current'' $s,d$ quark masses, appears in our
applications (see Sects.\ 4 and 5) when the  interference between the
(presumably) dominant LD contribution and the SD contribution is neglected.

It is difficult to estimate the coefficient $a_2(\mu^2)$ for $\mu\geq O(m_s)$
appearing in eq.~(5).  However, a smooth extrapolation from the
phenomenologically
obtained values $a_2(m_b^2) \simeq 0.24\pm 0.04$ and $a_2(m_c^2)=0.55 \pm
0.1$ [3,13] leads to $a_2(m^2_s)\geq 0.5$.

The couplings $g_{\psi_i}(m_{\psi_i}^2)$, $g_\rho(m^2_\rho), \
g_\omega(m^2_\rho)$ are readily obtained from leptonic decays of
these mesons, but their extrapolated values at
$q^2=0$ are  less trivial, especially for the $\psi_i$ states.  Photoproduction
data seems to indicate that $g_\rho^2(0) \simeq g^2(m^2_\rho)$, $g^2_\omega
(0)\simeq g^2(m^2_\omega)$ [17,18].  On the other hand, estimates in Ref.~[3]
using $\psi$ photoproduction data [18-20] give $g^2_\psi(0)=
(0.12\pm0.04)g^2_\psi(m^2_\psi)$.
In Ref.\ [3] it is also assumed that the same ratio holds for the excitations
$\psi^\prime, \ \psi^{\prime\prime}$, etc.

Making use of the above estimates as well as of the leptonic widths of the
relevant vector mesons [14] we obtain the numerical values
$g_\rho^2(0)/m^2_\rho \simeq 0.047$GeV$^2$,
$g_\omega^2(0)/m^2_\omega\simeq0.038$GeV$^2$ and
$\sum_i \frac{g^2_{\psi_i}(0)}{m^2_{\psi_i}} \simeq 0.041$GeV$^2$.  The
first two estimates should be accurate to about 10\% while the latter must
be considered only as a rough estimate, with an uncertainty of at least 40\%.
Once we derive the approximate relation (3) we will be able to give a far
more reliable estimate of $\sum_i \frac{g^2_{\psi_i}(0)}{m^2_{\psi_i}}$, which
is
consistent with the above central value.\\

\noindent{\bf 4. \ Application to the Decay $\Omega^-\rightarrow\Xi^-\gamma$
and Consequences}

We use the quark model of Ref.\ [10] to estimate the rate for the decay
$\Omega^-\rightarrow \Xi^-\gamma$, from the SD and LD contributions to
the $s\rightarrow d \gamma$ quark decay amplitude obtained in previous
sections.

For notational convenience, we define the constants $v\equiv|V_{cs}V^*_{cd}|
\simeq 0.22$ and $C_{\rm VMD} \equiv \left(\frac{2}{3} \sum_i
\frac{g^2_{\psi_i}(0)}{m^2_{\psi_i}} -  \frac{1}{2}
\frac{g^2_{\rho}(0)}{m^2_{\rho}} - \frac{1}{6}
\frac{g^2_{\omega}(0)}{m^2_{\omega}}\right)$.
The relative sign of the SD and LD contributions is determined by the theory
[3] so that
the full amplitude for the $s\rightarrow d \gamma$ transition can be
written as
\pagebreak

\begin{eqnarray}
A_{\rm TOT}(s\rightarrow d\gamma) &=& A_{\rm SD} + A_{LD} \nonumber \\ &=& -
\frac{eG_F}{\sqrt{2}} \bar{d}\sigma^{\mu\nu}
\left[\left(\frac{m_sF_2}{8\pi^2} +
\frac{va_2C_{\rm VMD}M_s}{M^2_s-M^2_d}\right)R \right. \nonumber \\&& +
\left. \left(\frac{m_dF_2}{8\pi^2} -
\frac{va_2C_{\rm VMD}M_d}{M^2_s-M^2_d}\right)L\right)s      \
F_{\mu\nu} \ .                 
\end{eqnarray}

Following Ref.\ [10] we then obtain
\begin{eqnarray}
\Gamma(\Omega^-\rightarrow\Xi^-\gamma) &=&
\frac{\alpha G^2_F}{12\pi^4}
\left(\frac{m_{\Xi^-}}{m_{\Omega^-}}\right) |\vec{q}|^3 \nonumber \\
&&\cdot \left\{ \left(m_sF_2 +
\frac{8\pi^2va_2C_{\rm VMD}M_s}{M^2_s-M^2_d}\right)^2 +
\left(m_dF_2 -
\frac{8\pi^2va_2C_{\rm VMD}M_d}{M^2_s-M^2_d}\right)^2 \right\} \ , \nonumber \\
\end{eqnarray}
where $\vec{q}$ is the photon momentum in the $\Omega^-$ rest frame and
the separate SD and LD contributions are exhibited explicitly.

In the absence of LD (VMD) contributions, we would obtain (for
$m_s\simeq175$MeV, $m_d \simeq 10$MeV, $F_2 \simeq 0.1$, see Sect.~2)
\begin{eqnarray}
\Gamma_{\rm SD}(\Omega^-\rightarrow\Xi^-\gamma)\simeq 6.4 \times 10^{-12}
\ {\rm eV}                   
\end{eqnarray}
which is far below the present experimental bound of
$\Gamma_{\exp}(\Omega^-\rightarrow\Xi^-\gamma)< 3.7 \times 10^{-9}$eV.
On the other hand, the large theoretical uncertainty of over 40\% in
the value of the sum $\sum_i\frac{g^2_{\psi_i}(0)}{m^2_{\psi_i}}$
(see Sect.~3) which appears in $C_{\rm VMD}$, would allow the LD contribution
to saturate this experimental bound.  In fact, the experimental limit
can be used to constrain $C_{\rm VMD}$ and hence
$\sum_i\frac{g^2_{\psi_i}(0)}{m^2_{\psi_i}}$.  Using typical values
$M_s\simeq0.5$GeV, $M_d\simeq 0.35$GeV for the constituent quark masses
and $a_2>0.5$ (see Sect.\ 3), we find
\begin{eqnarray}
\rule[-0.50cm]{0.02cm}{1.25cm}
C_{\rm VMD}
\rule[-0.50cm]{0.02cm}{1.25cm}
 =
\rule[-0.50cm]{0.02cm}{1.25cm}
\frac{2}{3} \sum_i
\frac{g^2_{\psi_i}(0)}{m^2_{\psi_i}}- \frac{1}{2}
\frac{g^2_{\rho}(0)}{m^2_{\rho}}- \frac{1}{6}
\frac{g^2_{\omega}(0)}{m^2_{\omega}}
\rule[-0.50cm]{0.02cm}{1.25cm}
< 0.01{\rm GeV}^2  \ .            
\end{eqnarray}
This constraint would be only slightly different, had we used current quark
mass parameters instead of $M_s$ and $M_d$.

The bound in eq.~(9) represents a remarkable cancellation at the 30\% level,
considering that
$\frac{1}{2} \frac{g^2_{\rho}(0)}{m^2_{\rho}}+ \frac{1}{6}
\frac{g^2_{\omega}(0)}{m^2_{\omega}}\simeq 0.030 {\rm GeV}^2$ (see Sect.\ 3).
We presume that this effect may stem from the combination of the GIM [21]
mechanism and the underlying SU(4)$_F$ symmetry, which if exact would
give a full cancellation (after inclusion of $\rho^\prime, \rho^{\prime\prime},
\dots$, $\omega^\prime, \ \omega^{\prime\prime}\dots$ states).  The
SU(4)$_F$ symmetry is known to be badly broken by the large mass of the
$c$ quark.  However, here we are comparing the form factors
$g^2_{\psi_i}(q^2)$,
$g^2_{\rho}(q^2)$,
$g^2_{\omega}(q^2)$ at a common scale $q^2=0$, which seems to ``restore''
this symmetry to some extent.  We have noticed that if
$|g_\phi(0)|\simeq|g_\phi(m_\phi^2)|$ [17,18], leading through $\phi$
leptonic width data [14] to $|g_\phi(0)|\simeq 0.24$GeV$^2$,
a completely analogous near cancellation occurs for the quantity
$C_{\rm VMD}^\prime \equiv -\frac{1}{3} \frac{g^2_{\phi}(0)}{m^2_{\phi}}
+ \frac{1}{2}
\frac{g^2_{\rho}(0)}{m^2_{\rho}}- \frac{1}{6}
\frac{g^2_{\omega}(0)}{m^2_{\omega}}$, which is relevant to LD effects in
$c\rightarrow u\gamma$ decay [6].  We obtain $C^\prime_{\rm VMD}\simeq -
1.8\times10^{-3}$GeV$^2$,
which represents a cancellation at a level better than 10\% for which
presumably the SU(3)$_F$ symmetry is responsible.

We note that the upper bound (9) on  $|C_{\rm VMD}|$ tells us that
although the LD effects are likely to dominate $s\rightarrow d\gamma$, they
can be at most a factor of about 25 larger than the SD contribution in the
amplitude.
This represents an
intermediate situation between the $b\rightarrow s\gamma$ decays where the
SD contribution clearly dominates [3,4,22] and the $c\rightarrow u\gamma$
decays where the SD effects are completely
negligible relative to the LD ones.\\

\noindent{\bf 5. \  Implications for the LD Contribution to $b\rightarrow
s\gamma$}

Because $\left(\frac{1}{2} \frac{g_\rho^2(0)}{m^2_\rho} +
\frac{1}{6} \frac{g_\omega^2(0)}{m^2_\omega}\right) \simeq 0.030$GeV$^2$,
eq.\ (9) implies that the approximate relation given in eq.~(3) must hold to
an accuracy of order 30\%.  This then independently determines
\begin{eqnarray}
\sum_i \frac{g^2_{\psi_i}(0)}{m^2_{\psi_i}} = 0.045\pm0.016{\rm GeV}^2 \ ,  
\end{eqnarray}
where our uncertainty in the values of $g_\rho(0)$ and $g_\omega(0)$ has been
folded in.  Notice that this result is in very good agreement with the
central value ($\simeq 0.041)$ estimated from $\psi$ photoproduction data in
Ref.\ (3), but the uncertainties there were larger (above 40\%).
Our results thus confirm previous assertions that the LD corrections are at the
few percent level only [3,4] and further show that these contributions are well
under control.  The amplitude for $b\rightarrow s\gamma$ including SD and LD
contributions can be expressed as [3]
\begin{eqnarray}
A_{\rm TOT}(b\rightarrow s\gamma)&=& - \frac{eG_F}{\sqrt{2}} V_{tb}V^*_{ts}
\left[\frac{1}{4\pi^2} m_bC_7^{\rm eff} (m_b) - a_2(m_b)
\frac{2}{3m_b} \sum_i \frac{g^2_{\psi_i}(0)}{m^2_{\psi_i}} \right] \nonumber \\
&\cdot & \bar{s}\sigma^{\mu\nu} RbF_{\mu\nu} \ .     
\end{eqnarray}
where $m_s, \ M_s$ have been neglected compared to $m_b$.  Using $a_2(m_b)
\simeq 0.24 \pm 0.04$[3], $C^{\rm eff}_7(m_b)=-0.30\pm0.03$ [16] and
$m_b = 4.8  \pm 0.2$GeV, we find that the LD contribution increases the
magnitude of the amplitude by $(4\pm2)\%$.\\

\noindent{\bf 6. Application to $B_s\rightarrow B_d^* \gamma$}

Another process where the $s\rightarrow d\gamma$ quark transition will
dominate is $B_s\rightarrow B^*_d\gamma$.  There are no pole contributions
and we explicitly estimated the LD contribution from light meson loops to
be smaller but comparable with the SD $s\rightarrow d\gamma$ contributions.
This is an unusual $B_s$ meson decay in the sense that it represents the
decay of the {\em light} quark in a $\bar{Q}q$ system.  Also, it has
a clear signature: two photons with energies of about 50MeV and 46MeV (the
second one coming from the decay $B_d^*\rightarrow B_d\gamma$), followed
by a usual $B_d$ decay.

We roughly estimate the $B_s\rightarrow B_d^*\gamma$ decay rate from
our $s\rightarrow d\gamma$ amplitude (6) by assuming that the spatial
wavefunctions
of the $s$ quark in the $B_s$ meson and the $d$ quark in the $B^*_d$ meson are
similar, and noting that the photon energy $(=50$MeV) is small compared to
the average momentum ($O$(700MeV)) of the light quark in  the bound state.
A ``free quark'' approximation should then give a reasonable estimate
of the transition  amplitude.  In terms of the effective $s\rightarrow d
\gamma$ Hamiltonian (6) we obtain for the decay rate:
\begin{eqnarray}
\Gamma(B_s\rightarrow B^*_d\gamma) &=&
\frac{\alpha}{16\pi^4} G_F^2|\vec{q}|^3
\left\{\left(m_sF_2 + \frac{8\pi^2va_2C_{\rm VMD}M_s}{M^2_s-M^2_d}\right)^2
\right.\nonumber \\
&& +
\left. \left(m_dF_2 - \frac{8\pi^2va_2C_{\rm VMD}M_d}{M^2_s-M^2_d}\right)^2
\right\}
\nonumber \\
\end{eqnarray}
where $\vec{q}$ is the photon momentum in the $B_s$ rest frame.  Comparing
to eq.~(7) and using the upper bound (1) we obtain
\begin{eqnarray}
\Gamma(B_s\rightarrow B^*_d\gamma) < 1.4 \times 10^{-20} \ {\rm GeV}  \ . 
\end{eqnarray}
Then, the present central value for the $B_s$ lifetime
$\tau_{B_s}\simeq 1.34 \times 10^{-12}s$ [14] gives a bound on the branching
ratio,
$BR(B_s\rightarrow B_d^*\gamma)< 3\times 10^{-8}$.  Although this is a very
rare decay mode, its unique signature and the large number of $B_s$ mesons
expected at $B$ meson factories and at
LHC-B $O(2\times10^{-11})$ [23] make it interesting.\\

\noindent{\bf 7. \ Conclusions}

Using a VMD approximation, we found that the LD contribution to the
$s\rightarrow d\gamma$ transition may be significantly larger than the SD
one, and could even lead to a saturation of the present experimental
upper limit on the decay rate for $\Omega^-\rightarrow \Xi^-\gamma$ (eq.\ (1)).
This result throws new light on this decay mode.  A further tightening of
this upper limit or a measurement of the $\Omega^-\rightarrow\Xi^-\gamma$ rate
would provide us with very useful information about the relative importance
of the LD and SD contributions to $s\rightarrow d\gamma$.  The present upper
bound already implies a non-trivial cancellation at a level of 30\% or
better in the LD contribution.  The resulting approximate relation (eq.~(3))
allowed
us to estimate the relative importance of the LD contribution to the
$b\rightarrow s\gamma$ transition amplitude.  Our estimate of $(4\pm
2)\%$ for this relative LD contribution agrees with earlier ones, which
had larger uncertainties.  Because the unusual process
$B_s\rightarrow B^*_d\gamma$ is also dominated by an $s\rightarrow d\gamma$
transition, its decay rate is related to that of $\Omega^-\rightarrow
\Xi^-\gamma$.  We find that a present limit on the latter (eq.\ (1)) implies
an upper bound for the branching ratio  $BR(B_s\rightarrow B^*_d\gamma)<3
\times 10^{-8}$, which is small but hopefully accessible
in future experiments.\\

\noindent{\bf Acknowledgements}

The research of A.I. and P.S. is supported in part by Grant 5421-2-95
from the Ministry of
Science and the Arts of Israel and by the Jubilee Fund of the Austrian
National Bank, Project 5051.  A.I. has been partially supported by the
Lady Davis Trustship.  The work of R.M. was supported in part by the
Natural Sciences and Engineering Research Council of Canada.  The work
of G.E. and P.S. has been supported in part by the Fund for Promotion of
Research
at the Technion and of G.E. by GIF.  We would like to thank\linebreak
 A. Buras for
useful correspondence.

\pagebreak

\noindent{\bf References}

\begin{enumerate}
\item B.A. Campbell and P.J. O'Donnell, Phys.\ Rev.\ {\bf D25}, 1989 (1982);
for review see: S. Pokorski, Invited talk at the Int.\ Europ.\ Conf.\ on
High Energy Physics, Marseilles, July 1993, Eds. J. Carr and M. Perrottet,
Edition Frontier (1994) p.~166.
\item R. Ammar et al., Phys.\ Rev.\ Lett.\ {\bf 71}, 674 (1993); M.S. Alam et
al.,
Phys.\ Rev.\ Lett.\ {\bf 74}, 2885 (1995).
\item N.G. Deshpande, X.-G. He and J. Trampetic, OITS-564-REV, HEP-PH/9412222.
\item E. Golowich and S. Pakvasa, Phys.\ Rev.\ {\bf D51}, 1215 (1995);
see however that according to the recent exclusive and inclusive analysis of
J. Soares in TRI-PP-95-6 (HEP-PH/9503285), uncertainties may still remain.
\item J.L. Hewett, SLAC-PUB-6521, HEP-PH/9406302.
\item G. Burdman, E. Golowich, J.L. Hewett and S. Pakvasa, SLAC-PUB-6692,
HEP-PH/9502329.
\item M.A. Shifman, A.I. Vainshtein and V.I. Zakharov,  Phys.\ Rev.\ {\bf D18},
2583 (1978); M. McGuigan and A.I. Sanda, Phys.\ Rev. {\bf D36}, 1413 (1987).
\item Ya.I. Kogan and M.A. Shifman, Sov.\ J.\ Nucl.\ Phys.\ {\bf 38}, 628
(1983).
\item S. Bertolini, M. Fabbrichesi and E. Gabrielli, Phys.\ Lett.\ {\bf B327},
1361 (1994).
\item  F. Gilman and M.B. Wise, Phys.\ Rev.\ {\bf D19}, 976 (1979);\\
R. Safadi and P. Singer, Phys.\ Rev.\ {\bf D37}, 697 (1988); {\bf D42},
1856 (E) 1990. Note that in the latter reference, a factor of  $\frac{1}{16}$
is missing in eqs.~(26) and (27).
\item I.F. Albuquerque et al., Phys.\ Rev.\ {\bf D50}, 18 (1994).
\item L. Bergstr\"{o}m and P. Singer, Phys.\ Lett.\ {\bf B169}, 297 (1986).
\item M. Baur, B. Stech and M. Wirbel, Zeit.\ f\"{u}r Physik {\bf C34} (1987).
\item Particle Data Group Phys.\ Rev.\ {\bf D50}, 1173 (1994).
\item P. Singer, Essays in honor of M. Roos (M. Chaichian and J. Maalapi,
eds.),
1991, p.~143.
\item A.J. Buras, M. Misiak, M. Munz and S. Pokorski Nucl.\ Phys.\ {\bf B424},
374 (1994).
\item K. Terasaki, Nuov.\ Cim.\ {\bf 66A}, 475 (1981).
\item E. Paul, Proceedings of the 1981 International Symposium on Lepton and
Photon Interactions at High Energies, Ed.\ W. Pfeil (Bonn, 1981) p.~301.
\item R.L. Anderson et al., Phys.\ Lett.\ {\bf B38}, 263 (1977).
\item J.J. Aubert et al., Phys.\ Lett.\ {\bf B89}, 267 (1980).
\item S.L. Glashow, J. Iliopoulos, L. Maiani, Phys.\ Rev.\ {\bf D2}, 1285
(1970).
\item M.R. Ahmady, D. Liu and Z. Tao, HEP-PH/9302209.  These authors pay
special attention to gauge invariance.  They obtain a large VMD contribution
to $b\rightarrow s\gamma$ from the $\psi$ states because of the assumption
$g^2_{\psi_i}(m^2_{\psi_i})^2\simeq g^2_{\psi_i}(0)$, which is a poor
approximation as estimated in ref. [3] and confirmed here (see eq.~(10)).
\item See e.g., W. Hoogland, invited talk presented at the 6th Intl. Symp.
Heavy
Flavour Physics, Pisa 6-9 June, 1995.
\end{enumerate}
\end{document}